# NON-LOCAL GRAPH-BASED PREDICTION FOR REVERSIBLE DATA HIDING IN IMAGES

*Qi Chang* [*], *Gene Cheung* [#], *Yao Zhao* [*], *Xiaolong Li* [*], *Rongrong Ni* [*]

[*] Institute of Information Science, Beijing Jiaotong University, [#] National Institute of Informatics

**ABSTRACT**

Reversible data hiding (RDH) is desirable in applications where both the hidden message and the cover medium need to be recovered without loss. Among many RDH approaches is prediction-error expansion (PEE), containing two steps: i) prediction of a target pixel value, and ii) embedding according to the value of prediction-error. In general, higher prediction performance leads to larger embedding capacity and/or lower signal distortion. Leveraging on recent advances in graph signal processing (GSP), we pose pixel prediction as a graph-signal restoration problem, where the appropriate edge weights of the underlying graph are computed using a similar patch searched in a semi-local neighborhood. Specifically, for each candidate patch, we first examine eigenvalues of its structure tensor to estimate its local smoothness. If sufficiently smooth, we pose a maximum a posteriori (MAP) problem using either a quadratic Laplacian regularizer or a graph total variation (GTV) term as signal prior. While the MAP problem using the first prior has a closed-form solution, we design an efficient algorithm for the second prior using alternating direction method of multipliers (ADMM) with nested proximal gradient descent. Experimental results show that with better quality GSP-based prediction, at low capacity the visual quality of the embedded image exceeds state-of-the-art methods noticeably.

## 1. INTRODUCTION

*Reversible data hiding* (RDH) denotes a class of techniques where both the embedded digital message and the cover medium can be recovered without loss at the decoder, with applications to copyright protection, secret communication, etc. Among many RDH approaches in the literature—*e.g.*, difference expansion (DE) [1–3], histogram shifting (HS) [4]—is *prediction-error expansion* (PEE) [5, 6], which exploits inherent smoothness in natural images for data embedding. PEE is the most popular approach at present. It consists of two steps: i) prediction of a target pixel value from context, and ii) embedding bits (expanding) or shifting according to the value of prediction-error. In general, higher prediction performance can lead to more embedded information bits (larger capacity), or lower signal distortion for the same target embedding capacity. Previous proposed prediction schemes include rhombus [6] and partial differential equations (PDE) [7], which are signal interpolation strategies assuming local smoothness.

In this paper, we focus on improving the prediction performance in PEE by leveraging on recent advance in graph-signal restoration [8–10]. Specifically, for each candidate pixel patch, we first examine eigenvalues of its structure tensor [11] to estimate its local smoothness. If it is sufficiently smooth, then assuming the self-similarity characteristic in natural images—-common in image denoising such as non-local means (NLM) [12]—we search for a similar patch in a semi-local neighborhood. The similar patch is used to compute appropriate edge weights for an underlying graph. We then pose pixel prediction as a *maximum a posteriori* (MAP) problem with suitable graph-signal smoothness priors: a quadratic graph Laplacian regularizer [9], or a graph total variation (GTV) term [13]. We show that the MAP problem with the first prior has a numerically stable closed-form solution. We then design an efficient algorithm for the second prior using *alternative direction method of multipliers* (ADMM) [14] with a nested proximal gradient descent [15].

Given a predicted value, following previous PEE schemes [5, 6] we embed an information bit as the difference between the predicted and the original pixel values. We show that this prediction structure can be executed in four individual layers in succession, increasing the overall embedding capacity. Experimental results show that with better quality graph-based prediction, at low capacity the visual quality of the embedded image exceeds state-of-the-art methods noticeably.

## 2. RELATED WORK

Tian [1] proposed a *difference expansion* (DE) method applying integer Haar wavelet transform to compute differences of pixel values. Then a message sequence is embedded into the vacancies of these differences via expanding. *Prediction-Error Expansion* (PEE) as an extension of DE is first proposed by Thodi and Rodriguez [5]. It exploits smoothness of natural images for prediction to improve embedding performance with minimal increase in distortion. *Histogram shifting* (HS), first proposed by Ni et al. [4], generates a histogram and shifts some bins to create space to embed bits. DE and HS are two most popular RDH technologies, and as a combination of PEE and HS, HS-PEE is the most commonly used method. It has two major steps: *prediction-error histogram* (PEH) generation and histogram modification. A PEH is generated by employing prediction to pixels. Because of the redundancy of naturals images, the histogram tends to obey a Laplacian-like distribution centered at or close to 0. A more accurate prediction scheme has a sharper histogram distribution centered at 0, resulting in less distortion. By expanding and shifting PEH, some close to 0 histogram bins are expanded to embed data, and other bins are shifted outwards to create space for expansion. At last, pixel values are modified to obtain the data embedded image. Following [5], many RDH techniques related to PEE have been proposed recently, some of them optimized the performance of prediction, and others designed effectively histogram modifying strategy [16–18].

Our method focuses on improving the prediction part with a conventional histogram modifying approach. The previous approaches are mostly based on predicted pixels' local context. Rhombus prediction with sorting is proposed in Sachnev et al. [6]. It uses a checkerboard and double-layered embedding strategy, and predicts pixel by calculating the mean of its four nearest neighboring pixels.

**Fig. 1**. Left: Piecewise smooth patch. Right: Smooth patch.

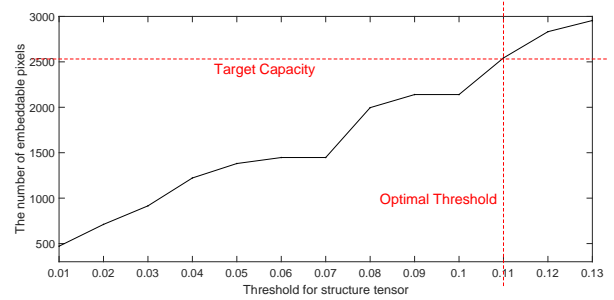

**Fig. 2**. Number of embeddable pixels versus threshold for eigenvalue $\lambda_{\min}$ of structure tensor for the first layer of `Airplane`.

Then all prediction errors are sorted by *local complexity* (LC): a candidate patch with smaller LC tends to have smaller prediction error and thus higher priority for data embedding. Ou et al. [7] proposed a *partial differential equations* (PDE) method. The initial prediction-error values are calculated by rhombus prediction. Then it is iteratively updated by considering the gradients of four directions based on local context. Dragoi et al. [19] proposed a local-prediction-based method which performs better than prior art such as *median edge detector* (MED) and *gradient-adjusted predictor* (GAP). It uses a least square predictor in a square block centered on the pixel without increasing any additional information. More recently, Chen et al. [20] proposed a directionally enclosed prediction and expansion (DEPE) method, observting that LC is not always proportional to the magnitude of prediction-error. It predicts a pixel using horizontal or vertical predictors, and only pixels where LC is proportional to the magnitude of prediction-error are used for embedding.

Different from these methods, our proposal leverages on recent image interpolation ideas such as NLM [12] and graph-signal restoration [8–10]. Other PEE-based schemes can also potentially benefit from adopting our prediction contribution.

## 3. SYSTEM OVERVIEW

### 3.1. Structure Tensor as Smoothness Criterion

We first overview our reversible data hiding system. Compared to previous PEE schemes, we mainly focus on improving pixel prediction accuracy, leveraging on recent advances in graph-based image restoration [8–10]. For natural images, a pixel in a smooth area—*e.g.*, the right block in Fig. 1—is more likely to be predictable locally, and thus has potential to embed an information bit. In previous prediction methods like [6], LC is computed for each candidate pixel, and only pixels with LC lower than a threshold (pre-selected to guarantee a target capacity) are predicted and expanded to embed a bit. However, for *piecewise smooth* (PWS) patch—*e.g.*, the left block in Fig. 1—LC is high and naïve local prediction methods like [6] cannot perform well.

Because our graph-based prediction scheme can also well predict pixels in PWS patches (to be detailed next), we propose a smoothness criterion based on *structure tensor* [11] that recognizes and permits PWS patches for embedding. Specifically, for each candidate pixel in each layer, we compute the eigenvalues of its structure tensor using the pixel's surrounding neighboring 8 pixels (to ensure reversibility). If the smaller of the two eigenvalues $\lambda_{\min}$ is smaller than a threshold $\tau$, we declare the patch as "predictable", *i.e.*, it is a valid candidate for embedding. Other candidate pixels that are not predictable are excluded. Eigenvectors of a structure tensor correspond to major and minor gradients of the patch, with eigenvalues representing the magnitude of the gradients. Thus a piecewise constant (PWC) patch has $\lambda_{\min} = 0$, since its minor gradient (direction parallel to its discontinuity) has zero magnitude.

We optimize $\tau$ to exclude unpredictable pixels and to ensure the number of predictable pixels reaches the target capacity for each layer. As an illustration, in Fig. 2 we observe that as threshold $\tau$ increases, the number of embeddable pixels in the first layer of image `Airplane` increases monotonically. We can thus estimate the optimal threshold $\tau^*$ via a simple binary search.

### 3.2. Semi-local Search for Similar Patch

For each remaining candidate pixel, assuming the self-similarity characteristic in natural images (as done in NLM for image denoising [12]), we search in a semi-local window to find the most similar patch (using its surrounding eight neighboring pixels) in terms of Euclidean distance of their mean-removed AC components. A patch that contains other to-be-embedded pixels will not be considered for matching. The best-matched patch is used to compute edge weights in a graph, then one of two graph-based prediction algorithms in Section 4 is performed.

### 3.3. Side Information

For natural images, the value of pixels should be within $[0, 255]$. *Location map* (LM) is commonly used to handle the underflow/overflow problem to ensure reversibility [1], which marks the locations of problematic pixels. In our approach, the maximum modification to each pixel is 1, and only boundary pixels will cause problems. Pixels with value 1 or 254 will be marked with 0 in LM. Pixels with value 0 (255) are modified to 1 (254). All these modified pixels will be marked with 1 in LM. Finally, the location map is lossless compressed and embedded into image.

To ensure reversibility, nine parameters—the message size, four thresholds for structure tensor of each layer and four compressed location map size—are embedded into the image. To reduce the side information size, the six parameters of last three layers are encoded as the difference with previous layers with one flag bit. For most images, the boundary pixels are just a few, so LM can be efficiently compressed. Specifically, the overhead is $6 + 9 + (1 + 7) \times 3 + 7 + (1 + 5) \times 3 = 64$ bits, which is comparable with existing PEE literature.

### 3.4. Multi-layer Bit Embedding

As shown in Fig. 4, we divide candidate pixels in an image into four non-overlapping sets, so that bits embedding can be executed in four layers in succession. Side informations are embedded into *Least Significant Bits* (LSBs) of the first line of image using LSB replacement. So, the embedded payload includes the replaced LSBs, the compressed LM and the message. For each layer, $1/4$ of the payload will be embedded, then a conventional PEE embedding mapping strategy as shown in Fig. 3 is used. Here we only consider the case that the maximum modification to each pixel value is restricted to 1, so only prediction error values 0 and -1 ('embeddable') can be expanded to embed bits as shown. Other prediction error values ('not embeddable') are shifted to ensure reversibility at the decoder (resulting in distortion). The pixels of the second layer are predicted and embedded after embedding in the first layer is completed, then prediction and embedding in the third and fourth layers in order.

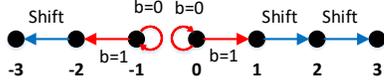

**Fig. 3**. Embedding strategy. When prediction-error is 0, it is expanded to 0 (or 1) with embedding bit 0 (or 1). When prediction-error is -1, it is expanded to -1 (or -2) with embedding bit 0 (or 1). Other prediction-errors shift outwards.

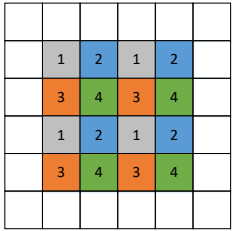

**Fig. 4**. Pixels of image are divided to four layer.

In the extraction stage, after extracting the side information, decoder uses the same non-local prediction strategy to predict the pixels of the fourth layer, extract embedded bits and restore original pixel values, then perform the same process for the third, second and first layers in order.

## 4. GRAPH-BASED PREDICTION

We propose two graph-based smoothness priors for pixel prediction in PEE: quadratic graph smoothness prior [9] and graph total variation (GTV) prior [13] previously used for image restoration, but for the first time in the literature we tailor their use for RDH. The first prior is computation-efficient, but the second can lead to better prediction performance in some cases.

### 4.1. Graph Construction from Similar Patch

We first construct an appropriate graph for a given target block $\mathbf{x} \in \mathbb{R}^N$. Denote by $\mathbf{H} \in \{0,1\}^{K \times N}$ a *sampling matrix* that selects $K$ observable pixels from a $N$-pixel patch $\mathbf{x} \in \mathbb{R}^N$ to an observation $\mathbf{y} \in \mathbb{R}^K$; *i.e.*, $\mathbf{y} = \mathbf{H}\mathbf{x}$. The remaining $N - K$ pixels will be predicted in our framework. Specifically for our embedding scheme, $\mathbf{x}$ is a $3 \times 3$ pixel patch, $\mathbf{y}$ is the surrounding eight pixels, and the center pixel is predicted.

Using $\mathbf{y}$, we first search for a similar patch $\mathbf{x}'$ in a defined semi-local neighborhood that is most similar to $\mathbf{x}$ in terms of Euclidean distance. Having found $\mathbf{x}'$, we construct a 8-connected similarity graph, where edge weight $w_{i,j}$ between neighboring pixels $i$ and $j$ is computed as follows:

$$w_{ij} = \exp\left(-\frac{\|\mathbf{l}_i - \mathbf{l}_j\|_2^2}{\sigma_l^2} - \frac{\|x_i' - x_j'\|_2^2}{\sigma_x^2}\right) \quad (1)$$

where $\mathbf{l}_i$ and $x_i'$ are location and pixel intensity of pixel $i$, and $\sigma_l$ and $\sigma_x$ are two parameters. We can then define an *adjacency matrix* $\mathbf{W}$ where $W_{i,j} = w_{i,j}$, a diagonal *degree matrix* $\mathbf{D}$ where $D_{i,i} = \sum_j W_{i,j}$, and a *combinatorial graph Laplacian matrix* $\mathbf{L} = \mathbf{D} - \mathbf{W}$.

### 4.2. Quadratic Graph Smoothness Prior

Given the graph Laplacian $\mathbf{L}$, we can formulate a *maximum a posteriori* (MAP) problem as follows. Using a graph Laplacian regularizer [9] as a signal prior, we can write the following objective:

$$\min_{\mathbf{x}} \|\mathbf{y} - \mathbf{H}\mathbf{x}\|_2^2 + \gamma \mathbf{x}^T \mathbf{L} \mathbf{x} \quad (2)$$

where $\gamma > 0$ is a parameter to trade off the fidelity term (negative log likelihood term in Bayesian terminology) and the prior term.

(2) is a linear combination of two quadratic terms in optimization variable $\mathbf{x}$. Hence to optimize (2), we take the derivative of (2) with respect to $\mathbf{x}$, equate it to 0 and solve for the closed form solution $\mathbf{x}^*$:

$$\mathbf{x}^* = (\mathbf{H}^T \mathbf{H} + \gamma \mathbf{L})^{-1} \mathbf{H}^T \mathbf{y} \quad (3)$$

We show that the matrix $\mathbf{H}^T \mathbf{H} + \gamma \mathbf{L}$ in our case is invertible. Assuming positive edge weights, one can show that $\mathbf{L}$ is positive semi-definite (PSD), and has (unnormalized) constant vector $\mathbf{1}$ as eigenvector corresponding to eigenvalue 0. Because $\mathbf{H}$ is a sampling matrix containing only non-negative entries, $\mathbf{H}\mathbf{1}$ has entries strictly greater than 0, and $\mathbf{1}$ cannot be an eigenvector corresponding to eigenvalue 0 for $\mathbf{H}^T \mathbf{H}$, which is PSD. Hence there does not exist a vector $\mathbf{v}$ such that $\mathbf{v}^T (\mathbf{H}^T \mathbf{H} + \gamma \mathbf{L}) \mathbf{v} = 0$. Since $\mathbf{H}^T \mathbf{H} + \gamma \mathbf{L}$ is at least PSD by Weyl's inequality but does not contain eigenvalue 0, it is positive definite (PD) and thus invertible.

### 4.3. Graph Total Variation Prior

The second prior is GTV [13], resulting in a $l_2$ / weighted-$l_1$ optimization problem:

$$\min_{\mathbf{x}} \|\mathbf{y} - \mathbf{H}\mathbf{x}\|_2^2 + \gamma \sum_{i,j} w_{i,j} |x_i - x_j| \quad (4)$$

where $\gamma$ again is a tradeoff parameter. To solve (4), we first rewrite it as follows. We first define $z_{i,j} = x_i - x_j$. Then (4) becomes,

$$\min_{\mathbf{x},\mathbf{z}} \|\mathbf{y} - \mathbf{H}\mathbf{x}\|_2^2 + \gamma \sum_{i,j} w_{i,j} |z_{i,j}|$$
$$\text{s.t.} \quad z_{i,j} = x_i - x_j, \ \forall (i,j) \in \mathcal{E} \quad (5)$$

To solve (5), unlike [13] that employed a primal-dual algorithm for an unconstrained GTV objective, we design a new algorithm based on *Alternating Direction Method of Multipliers* (ADMM) [14] with a nested proximal gradient descent [15]. We first write the set of linear constraints for connected pixel pairs in matrix form:

$$\mathbf{z} = \mathbf{F}\mathbf{x} \quad (6)$$

where $\mathbf{z} \in \mathbb{R}^M$ and $\mathbf{F} \in \{-1, 0, 1\}^{M \times N}$. Specifically, for each $z_{i,j}$, the corresponding row in $\mathbf{F}$ has all zeros except entries $i$ and $j$ that have 1 and $-1$ respectively. We can now rewrite (5) in ADMM scaled form as follows:

$$\min_{\mathbf{x},\mathbf{z}} \|\mathbf{y} - \mathbf{H}\mathbf{x}\|_2^2 + \gamma \sum_{i,j} w_{i,j} |z_{i,j}| + \frac{\rho}{2} \|\mathbf{F}\mathbf{x} - \mathbf{z} + \mathbf{u}\|_2^2 + \text{const} \quad (7)$$

where $\rho > 0$ is a Lagrange multiplier. As typically done in ADMM, we solve (7) by iteratively minimizing $\mathbf{x}$ and $\mathbf{z}$ and updating $\mathbf{u}$ one at a time in turn until convergence as follows.

*4.3.1.* **x**-*minimization*

To minimize $\mathbf{x}$ having $\mathbf{z}^k$ and $\mathbf{u}^k$ fixed, we take the derivative of (7) with respect to $\mathbf{x}$ and set it to 0:

$$\left(2\mathbf{H}^T\mathbf{H} + \rho\mathbf{F}^T\mathbf{F}\right)\mathbf{x}^* = 2\mathbf{H}^T\mathbf{y} - \rho\mathbf{F}^T(\mathbf{u}^k - \mathbf{z}^k) \qquad (8)$$

Because $\mathbf{F}$ is an inter-pixel difference operator, it is easy to see that $\mathbf{F}\mathbf{1} = \mathbf{0}$, and $\mathbf{1}$ is an eigenvector corresponding to eigenvalue 0 for $\mathbf{F}^T\mathbf{F}$. However, using the same previous reasoning, $\mathbf{1}$ cannot be an eigenvector corresponding to eigenvalue 0 for $\mathbf{H}^T\mathbf{H}$. Given $\mathbf{H}^T\mathbf{H}$ and $\mathbf{F}^T\mathbf{F}$ are both PSD, there does not exist a vector $\mathbf{v}$ such that $\mathbf{v}^T(2\mathbf{H}^T\mathbf{H} + \rho\mathbf{F}^T\mathbf{F})\mathbf{v} = 0$. Since $2\mathbf{H}^T\mathbf{H} + \rho\mathbf{F}^T\mathbf{F}$ must be at least PSD (again by Weyl's inequality) but has no eigenvalue 0, we can conclude that it is PD and invertible. Thus $\mathbf{x}^*$ can now be readily computed as a full-rank system of linear equations in (8).

*4.3.2.* **z**-*minimization*

Keeping $\mathbf{x}^{k+1}$ and $\mathbf{u}^k$ fixed, the optimization for $\mathbf{z}$ becomes:

$$\min_{\mathbf{z}} \frac{\rho}{2}\|\mathbf{F}\mathbf{x}^{k+1} - \mathbf{z} + \mathbf{u}^k\|_2^2 + \gamma\sum_{i,j}w_{i,j}|z_{i,j}| \qquad (9)$$

where the first term is convex and differentiable, and the second term is convex but non-differentiable. We can thus use *proximal gradient* [15] to solve (9). The first term has gradient $\nabla_{\mathbf{z}}$:

$$\nabla_{\mathbf{z}}(\mathbf{x}^{k+1}, \mathbf{z}, \mathbf{u}^k) = -\rho(\mathbf{F}\mathbf{x}^{k+1} - \mathbf{z} + \mathbf{u}^k) \qquad (10)$$

We can now define a proximal mapping $\text{prox}_{g,t}(\mathbf{z})$ for a convex, non-differentiable function $g(\ )$ with step size $t$ as:

$$\text{prox}_{g,t}(\mathbf{z}) = \arg\min_{\boldsymbol{\theta}}\left\{g(\boldsymbol{\theta}) + \frac{1}{2t}\|\boldsymbol{\theta} - \mathbf{z}\|_2^2\right\} \qquad (11)$$

We know that for our weighted $l_1$-norm in (9), the proximal mapping is just a soft thresholding function:

$$\text{prox}_{g,t}(z_{i,j}) = \begin{cases} z_{i,j} - t\gamma w_{i,j} & \text{if } z_{i,j} > t\gamma w_{i,j} \\ 0 & \text{if } |z_{i,j}| \leq t\gamma w_{i,j} \\ z_{i,j} + t\gamma w_{i,j} & \text{if } z_{i,j} < -t\gamma w_{i,j} \end{cases} \qquad (12)$$

We can now update $\mathbf{z}^{k+1}$ as:

$$\mathbf{z}^{k+1} = \text{prox}_{g,t}(\mathbf{z}^k - t\nabla_{\mathbf{z}}(\mathbf{x}^{k+1}, \mathbf{z}^k, \mathbf{u}^k)) \qquad (13)$$

We compute (13) iteratively until convergence.

*4.3.3.* **u**-*update*

Finally, we can update $\mathbf{u}^{k+1}$ simply:

$$\mathbf{u}^{k+1} = \mathbf{u}^k + (\mathbf{F}\mathbf{x}^{k+1} - \mathbf{z}^{k+1}) \qquad (14)$$

$\mathbf{x}$, $\mathbf{z}$ and $\mathbf{u}$ are iteratively optimized in turn using (8), (13) and (14) until convergence.

## 5. EXPERIMENTS

We compare our porposed data embedding scheme with two state-of-the-art methods, Sachnev et al. [6] and Ou et al. [7], in terms of capacity-distortion performance. We chose these two comparison methods because they both focus on improvement in the prediction and sorting components without changes to the histogram modifying component.

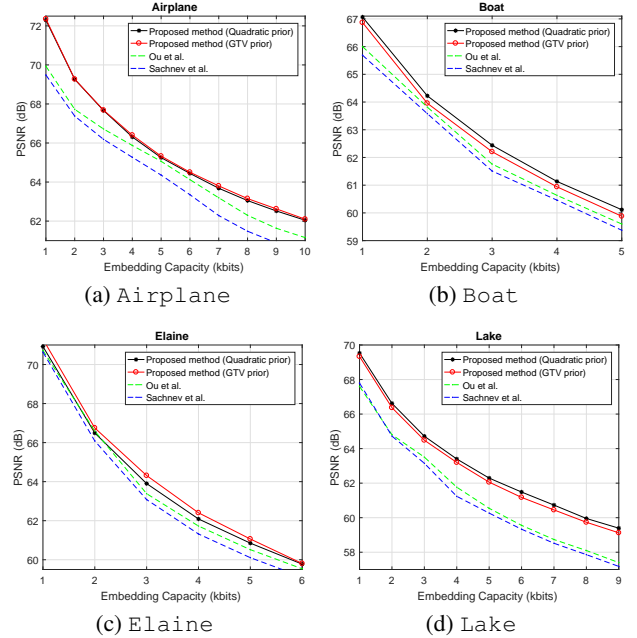

**Fig. 5**. (a)–(d): PSNR versus embedding capacity for `Airplane`, `Boat`, `Elaine` and `Lake` using our proposed method, Ou *et al.*[7] and Sachnev *et al.* [6].

Four standard $512 \times 512$ sized gray-scale images are used for testing: `Airplane`, `Boat`, `Elaine` and `Lake`. We focus the comparison on low capacity, since using semi-local search for similar patches inevitably leads to the limited capacity. As shown in Fig. 5(a) through (d), our proposal using the quadratic graph prior has up to 2.81dB and 2.36dB gains in PSNR over [6] and [7], respectively. Here, the selected parameters are $\sigma_l = \sigma_x = 0.5$, $\gamma = 0.5$, $\rho = 5$, $t = 0.1$. The size of the semi-local neighborhood used to find a similar patch is $31 \times 31$. Different thresholds $\tau$ are optimized for different layers for different images, where the range is $[0, 5]$. As shown in Fig. 2, when $\tau$ is 0.11, it is sufficient to meet the 10000 capacity requirement for the first layer of image `Airplane` when employing quadratic prior. For the rest of the layers, $\tau$ are 0.14, 0.16, 0.18 respectively. For image `Airplane` and `Elaine`, GTV prior has a maximum gain of 0.12dB and 0.41dB in PSNR over the quadratic graph prior in low capacity. In other regions, the two priors are comparable.

## 6. CONCLUSION

Prediction-error expansion (PEE) is a reversible data hiding (RDH) approach with two steps: i) prediction of a target pixel value based on local context; and ii) embedding bits (expanding) or shifting according to the value of prediction-error. In this paper, we improve the pixel prediction performance using one of two proposed graph-signal smoothnes priors: graph Laplacian regularizer, and graph total variation (GTV). While the posed inverse problem using the first prior has a closed-form solution, we design an algorithm for the second prior using alternating direction method of multipliers (ADMM) with nested proximal gradient descent. Our embedding scheme can be executed in four individual layers in succession, resulting higher embedding capacity. Experimental results show that with better graph-based prediction, visual quality of the embedded images exceeds state-of-the-art methods.